\documentclass[
 reprint,
 amsmath,amssymb,
 aps,
pra,
]{revtex4-2}

\usepackage{graphicx}
\usepackage{dcolumn}
\usepackage{bm}
\usepackage[english]{babel}
\begin{document}

\title{Cluster States Generation with a Quantum Metasurface}

\author{Yehonatan Levin}
 \email{yehonatan.levin@mail.huji.ac.il}
\author{Uri Israeli}
\author{Rivka Bekenstein}
\affiliation{
 Racah Institute of Physics, Hebrew University of Jerusalem, Jerusalem 91904, Israel
}

\begin{abstract}
We investigate the implementation of photonic cluster state generation protocols using quantum metasurfaces comprising sub-wavelength atomic arrays which enables quantum-controlled reflectivity. These cluster states are generated using fundamental quantum logic gates and enable wide-ranging applications in quantum computation and communication. In the past few years, certain protocols have been developed, but their physical realizations induces natural losses on the system mainly originated from coupling the photonic structures, setting a limit on the efficiency and maximal qubit number. In this paper, we examine a physical implementation of two specific protocols for generating distinct cluster states: a two-dimensional cluster state and a tree cluster state. Our approach leverages the unique properties of a quantum metasurface and its free space settings to implement two-qubit quantum-logic gates, namely CNOT, CZ, and E gates, with practical fidelities exceeding 0.9, and potential speed-up due to parallelism. In addition, we analyze these protocols fidelities for practical conditions of potential implementation experiments, such as thermal fluctuation of trapped atoms.
\end{abstract}
\maketitle
\section{Introduction}

While the scientific community is focused on implementing quantum applications, any quantum information processing task requires states of many entangled qubits. A useful resource for quantum information processing are highly entangled qubit states -- cluster states. Cluster states serve as valuable resources for one-way quantum computation and quantum communication purposes. In the former, cluster states are employed as a basis for measurement-based quantum computation \cite{briegel_measurement-based_2009, raussendorf_one-way_2001}, where in the latter, specific entangled states facilitate error correction and enable quantum encoding for distributed communication \cite{borregaard_one-way_2020, gisin_quantum_2002}.

Specifically, photonic qubits and optical systems offer several advantages for cluster state generation, due to their resilience at room temperature and the absence of interactions among themselves. Conventional methods for generating photonic cluster states involve cavity quantum electrodynamics (QED) systems \cite{wilk_single-atom_2007, reiserer_quantum_2014, zou_generating_2005, hilaire_near-deterministic_2023}. This method has natural loss mechanisms as a photon has to enter and leave the cavity, which imposes scattering losses. Other system of metasurfaces presented manipulation of the entangled degrees of freedom of single photons \cite{stav_quantum_2018, devlin_arbitrary_2017}, and metasurfaces nonlinear response enabled emission of entangled photons \cite
{ma_engineering_2024, santiago-cruz_photon_2021, santiago-cruz_resonant_2022, zhang_spatially_2022, ma_quantum_2025, jia_polarization-entangled_2025}. However, this method relies on post-selection and limited in its efficiency. 

A newly developed system of Quantum Metasurfaces (QMs) has emerged as a promising tool for achieving high-fidelity operations between atoms and photons \cite{bekenstein_quantum_2020}. This system is based on the enhanced response of sub-wavelength atom arrays to light due to photon-exchange interaction \cite{shahmoon_cooperative_2017, asenjo-garcia_exponential_2017, bettles_enhanced_2016} along with quantum control of single atoms. By quantum controlling the state of a single atom the array can switch from transmissive to reflective, enabling quantum control over the photon state by scattering.

\begin{figure}[ht!]
\centering\includegraphics[width=\columnwidth]{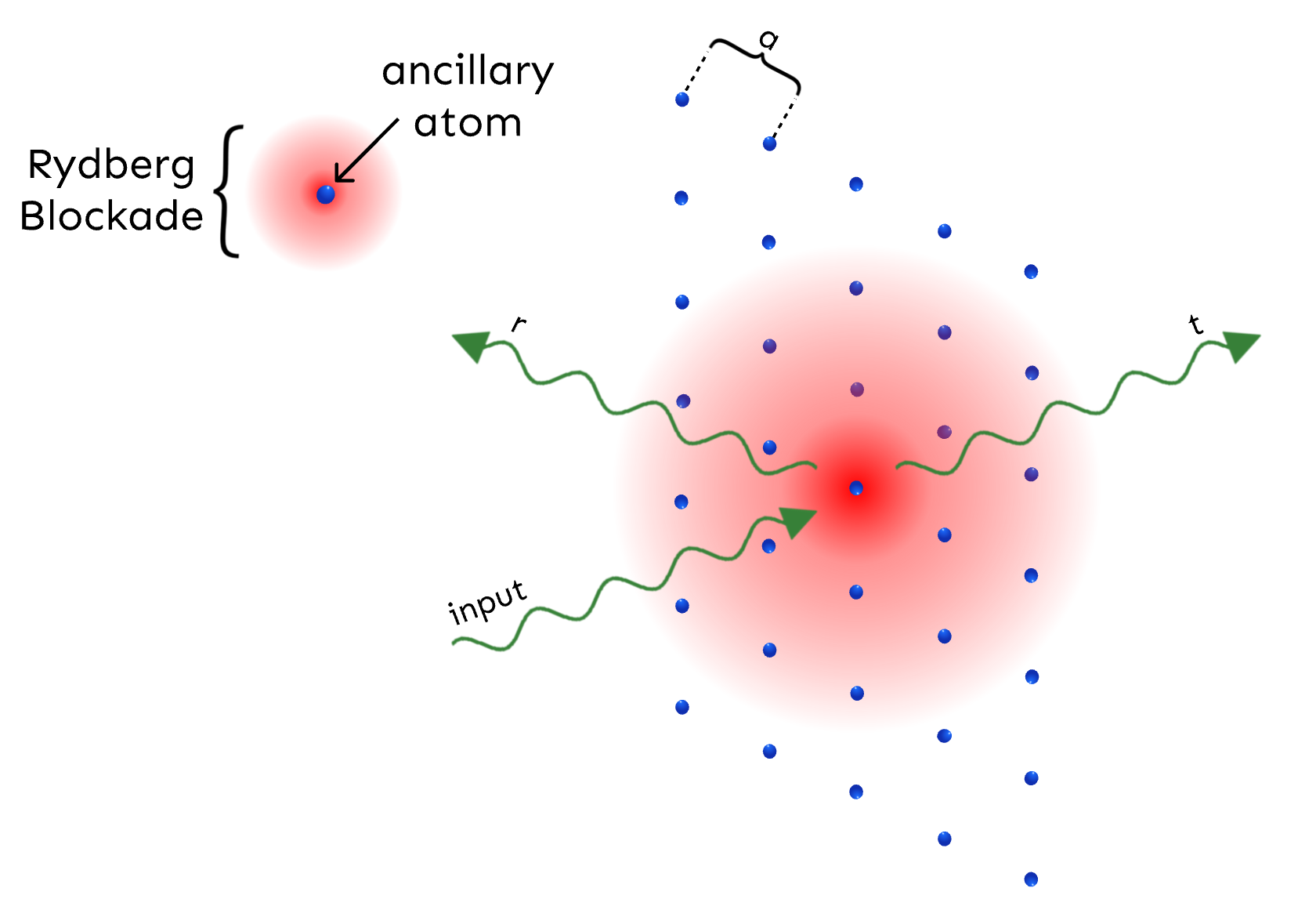}
\caption{A Quantum Metasurface featuring an ancillary atom at its center. $a$ is the lattice constant. Upon superposition of the ancillary atom $\frac{1}{\sqrt{2}} \left(\left|g\right\rangle + \left|r\right\rangle\right)$, an incoming photon with left-handed circular polarization $\left|0\right\rangle_p$ transitions into a superposition of both reflection and transmission $\frac{1}{\sqrt{2}} \left( \left|g\right\rangle \left|0\right\rangle_p + \left|r\right\rangle \left|1\right\rangle_p \right)
$, where the reflected photon was flipped to right-handed circular polarization $\left|1\right\rangle_p$, for ideal transmission and reflection coefficients.}
\end{figure}

Recent experimental progress has demonstrated the feasibility of QMs \cite{rui_subradiant_2020, srakaew_subwavelength_2023}. In addition, some works develop ways to entangle atomic qubits within sub-wavelength atomic arrays \cite{antman_ron_atom-atom_2024, shah_quantum_2024, patti_controlling_2021}. The quantum metasurface system manifest a deterministic preparation of photonic states based on scattering events (non post-selection), and naturally resides in a free-space setting, hence minimizing scattering losses. Some efforts are aimed to achieve entanglement in photonic free-space settings with atomic arrays \cite{PhysRevA.98.043825, PhysRevResearch.7.L022014}.  While this analysis of QM initiated various works exploring nonlinear optical processes, and the behavior of more complex geometrical configurations \cite{fernandez-fernandez_tunable_2022, masson_universality_2022, rubies-bigorda_photon_2022, shah_quantum_2024}, there is still a need to examine their effectiveness in generating photonic cluster states.

In this work we do exactly that, we develop protocols for generating photonic cluster states with quantum metasurfaces. In particular, we propose implementations of two distinct protocols for generating cluster states using a single quantum metasurface. These protocols enable high-fidelity gates between an ancillary atom and photons, eliminating the need for a cavity setup. Importantly, our approach allows for the production of scalable cluster states without requiring scaling the entire system by employing sequential photons. Additionally, we harness the spatial degree of freedom provided by the QM to simultaneously control multiple photonic qubits, offering potential speed-up of quantum operations. The set of quantum gates form versatile building blocks for various quantum operations, as detailed below.

\begin{figure*}[ht!]
\centering
\includegraphics[width=0.8\textwidth,height=0.45\textheight]{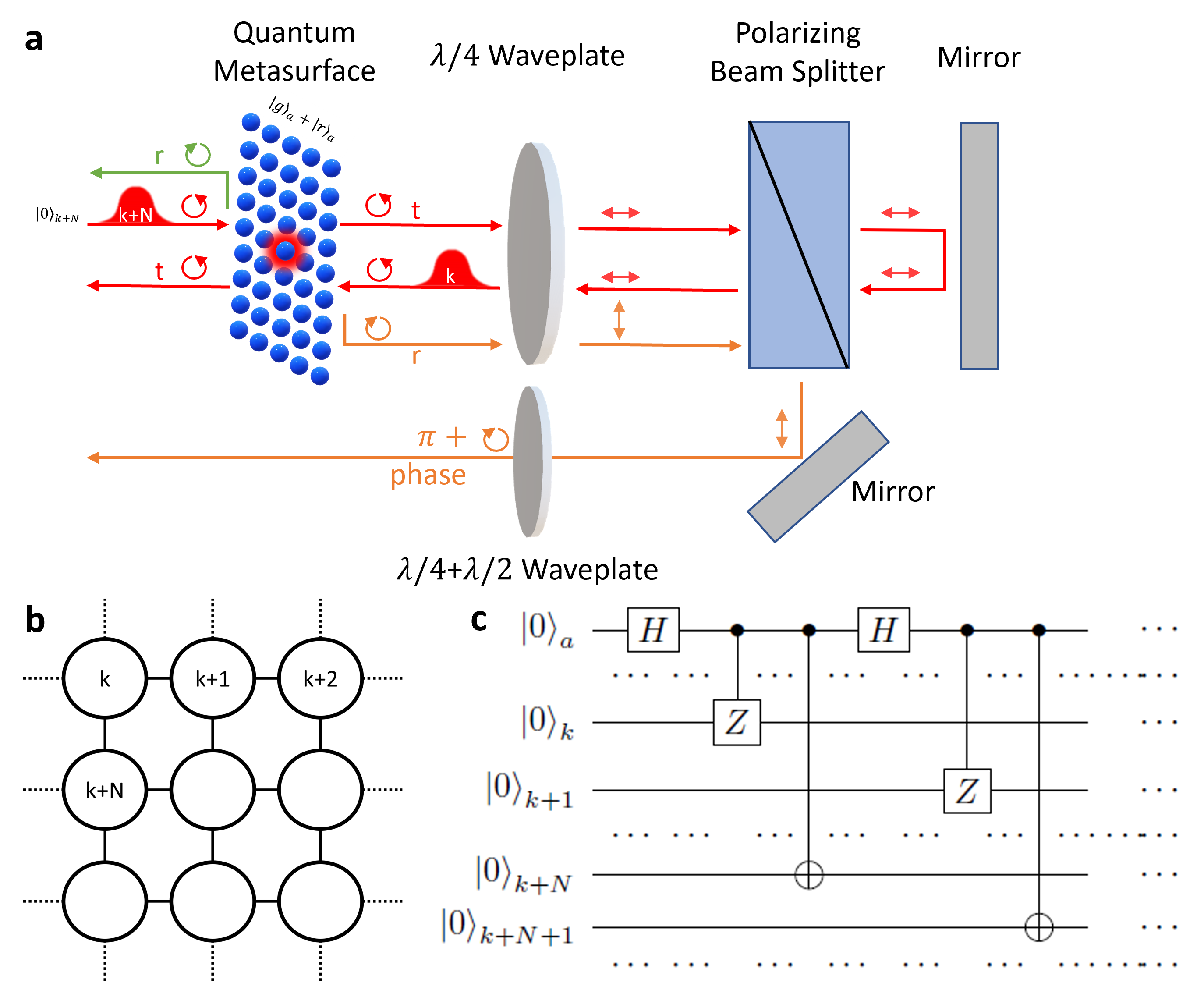}
\caption{\label{2D system}\textbf{a}\label{fig:2D}. Scheme for generating a scalable 2D cluster state. Sequential application of CNOT, CZ, and Hadamard gates with the Quantum Metasurface (QM) as control qubit. The incident photon (red) interacts with the QM, which is initialized in a superposition state, resulting in a superposition of reflection (green) and transmission (red). After passing through a wave-plate and a polarizing beam splitter (PBS), the photon returns to the QM, which remains in a superposition state. The reflected portion of the photon is redirected by the PBS (orange) due to its orthogonal polarization. The orange path has a different length, resulting a desired phase difference.
\textbf{b}. Scheme of a part of a 2D cluster state of width $N$. Qubit k is an arbitrary qubit inside the grid entangled to its four nearest neighbors.
\textbf{c}. Quantum circuit representation equivalent to the generation of four nodes out of the 2D cluster state. Each node $k$ is being connected to its $k+1$ neighbor and to its $k+N$ neighbor using CNOT and CZ gate with the ancilla.}
\end{figure*}

\subsection*{Quantum metasurface within an optical path for logic control of photon state}

To realize these protocols, we employ a quantum metasurface integrated into an optical setup. Our system consists of a quantum metasurface; a sub-wavelength atomic array of three-level atoms that facilitate coherent atom-photon interactions. The array comprise three-level atoms with ground ($\left|g\right\rangle$), excited ($\left|e\right\rangle$), and Rydberg ($\left|r\right\rangle$) states, arranged in a lattice where long-range dipole-dipole interactions enable collective responses to incident light. A pivotal feature is electromagnetically induced transparency (EIT), achieved by coupling a weak probe field resonant with the ($\left|g\right\rangle \leftrightarrow\left|e\right\rangle$) transition and a strong control field resonant with ($\left|e\right\rangle \leftrightarrow\left|r\right\rangle$). Under EIT conditions, the array becomes transparent to the probe, allowing unimpeded transmission; however, exciting ancillary atoms to Rydberg states disrupts EIT via level shifts, creating reflective regions that modulate the photonic amplitude. 

This disruption is governed by the Rydberg blockade, wherein excitation of one atom induces strong van der Waals or dipole-dipole interactions that shift nearby atoms' energy levels, suppressing further excitations within a blockade radius $R_b$. In this architecture, $R_b$ sets the spatial extent of coherent control, facilitating EIT switching and entanglement generation across the array. By preparing superpositions of $\left|g\right\rangle + \left|r\right\rangle$ for a single atom, the array realizes a superposition of reflective and transmissive states.   
The system analyzed in this work has architecture that incorporates a quantum metasurface together with optical elements, to generate the specific phase of part of the photon state by delaying it relative to other parts. For this goal we utilize optical elements such as wave plates, polarizing beam splitters, and mirrors, configured along designated propagation paths, as described below and depicted in Figures \ref{2D system} and \ref{tree}.

\section{Protocols for generation of highly entangled states}
\subsection{Two-dimensional cluster state generation}
We first show how the quantum metasurface can be utilized to generate two-dimensional (2D) cluster state, which are promising for measurement-based quantum computations \cite{briegel_measurement-based_2009}. To make these clusters a viable option we would like to analyze a scalable system that enables many-qubit states. Our protocol requires a sequential application of CZ, CNOT, and Hadamard gates \cite{pichler_universal_2017}. This sequence entangles qubits systematically to form the 2D cluster structure. Specifically, a CZ gate is applied between the ancilla and the $k^{th}$ qubit, followed by a CNOT gate between the ancilla and the $k^{th}+N$ qubit. Finally, a Hadamard rotation is performed on the ancilla. The quantum circuit depicted in Figure \ref{fig:2D}c illustrates the overall scheme, combining these gates as explained above. 

\begin{figure*}[ht!]
\centering
\includegraphics[width=0.9\textwidth,height=0.45\textheight]{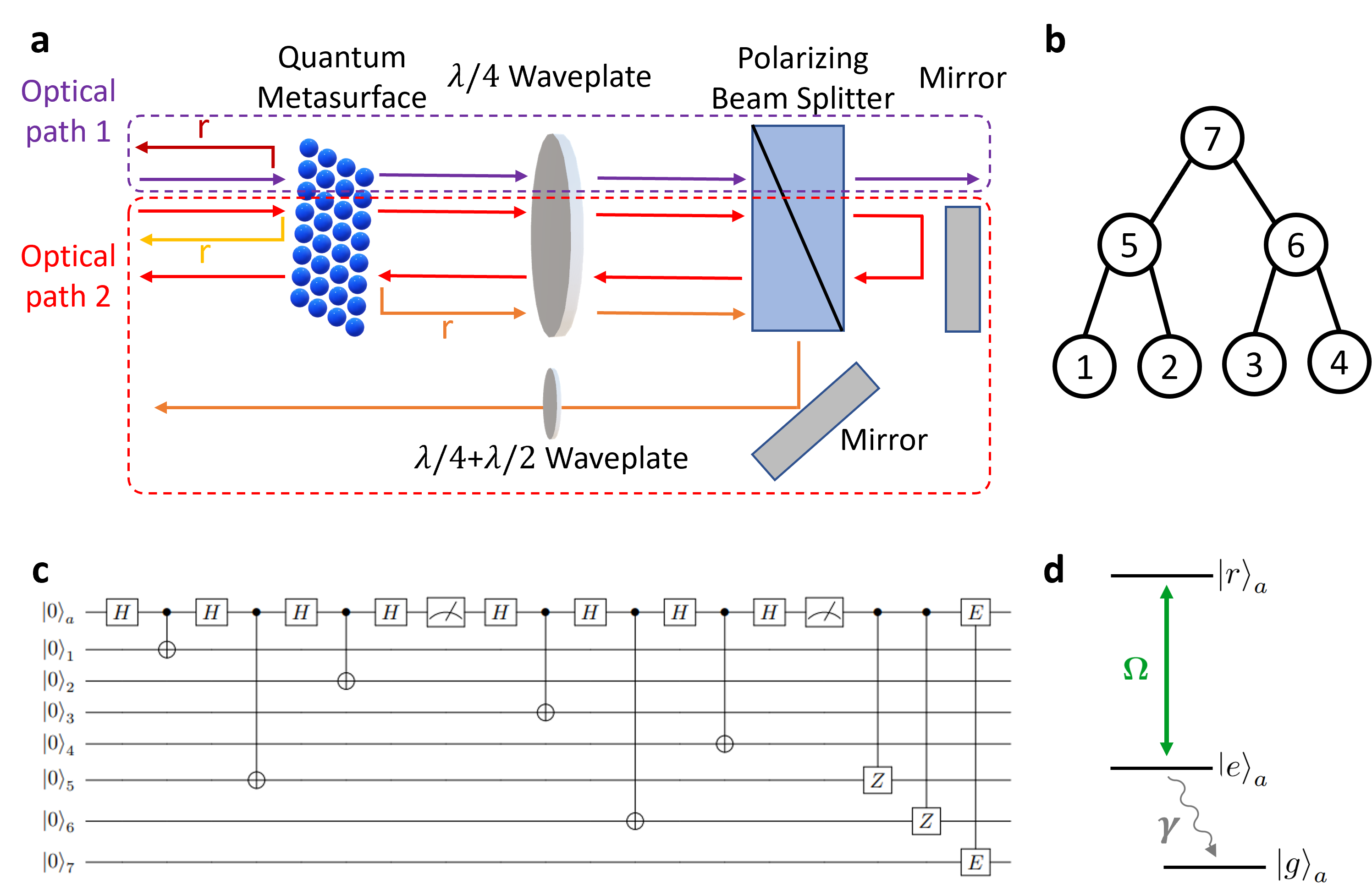}
\caption{\label{tree}\textbf{a}. Diagram illustrating the Tree cluster creation protocol. Two optical axes are depicted: Optical path 1 implements a CNOT gate as part of the E gate, where the incident photon (purple) interacts
with the QM, which is initialized in a superposition state, resulting in a superposition of reflection (dark red)
and transmission (purple).  Optical path 2 is identical to the diagram in Figure \ref{2D system}a, and accommodates photonic qubits 5 and 6, enabling the creation of a CZ gate for the necessary entanglement, following a similar procedure as in Figure \ref{2D system}a. 
\textbf{b}. Scheme of a binary tree with a height of three, which can be employed for information recovery in quantum communication.
\textbf{c}. Quantum circuit representation equivalent to the generation of a three-level binary tree. Both sub-trees are constructed independently, followed by the entanglement of their roots (nodes 5,6, see d) with the ancilla. Subsequently, the ancilla's state is transferred to another photon via the E gate, culminating in the tree's formation.
\textbf{d}. Implementation of the E gate, using atomic levels of the ancilla. After a CNOT gate, a $\pi$ pulse is applied, causing the state \(\left|r\right\rangle\) to transition to \(\left|e\right\rangle\). This pulse selectively drives the \(\left|e\right\rangle \leftrightarrow\left|r\right\rangle\)
transition due to resonance, while leaving \(\left|g\right\rangle\) unaffected because the pulse frequency is far detuned from any transition involving \(\left|g\right\rangle\). Subsequently, by observing the population in the state \(\left|g\right\rangle\)\ after a decay process, the success of the gate operation is verified, regardless of the initial state. }
\end{figure*}

\label{CNOT}
We now describe the specific implementation with a QM.
The protocol requires CNOT, CZ between ancilla and photon, and a Hadamard gate for the ancilla. The CNOT gate emerges as the most intuitive gate to construct using the QM, leveraging the properties of an ancillary atom within the metasurface as the control qubit. The ancilla atom's basis is the ground and Rydberg states, $\{\left|r\right\rangle, \left|g\right\rangle\}$, while the photon's basis is the circular polarization basis $\{\left|0\right\rangle_p, \left|1\right\rangle_p\}$. When the ancilla atom is in the ground state, a scattering event preserves both the atom and the photon states. In contrast, when the atom occupies the Rydberg state, right circular polarization converts to left circular polarization and vice versa, culminating in the formation of a CNOT gate.

For the implementation of the CZ gate, we designed the optical path depicted in Figure \ref{2D system}a. The photon is engineered so that the state \(\left|r\right\rangle_a\left|1\right\rangle_p\) — representing the ancilla on the  QM in the Rydberg state and the photon in a right-handed circular polarization, respectively — travels a unique optical path. The other states—\(\left|r\right\rangle_a\left|0\right\rangle_p\), where the ancilla is in the Rydberg state and the photon in a left-handed circular polarization, and \(\left|g\right\rangle_a\left|0\right\rangle_p, \left|g\right\rangle_a\left|1\right\rangle_p\), where the ancilla is in ground state and the photon is in either circular polarization—travel along the same optical path. The differential length of this path is carefully selected to impart a relative $\pi$ phase on the photon.

Formally, we prepare the state \[\left|\psi_{2D}\right\rangle = \left(\prod_{i=1}^{N} CNOT_{a,i+N} CZ_{a,i} H_{a}\right) \left|g\right\rangle_a \underset{i}{\bigotimes} \left|0\right\rangle_i.\] 
In each step, we apply a Hadamard gate to the ancilla atom (which acts as the control qubit) followed by controlled-Z and controlled-NOT gates on the photonic qubits as target qubits. This sequential interaction entangles the ancilla with two photonic qubits at each step, effectively generating a highly-entangled state of many photonic qubits (1000s of photonic qubits for practical Rubidium 70S Rydberg state, with collective linewidth of about $72Mhz$ \cite{srakaew_subwavelength_2022, levine_high-fidelity_2018}).

\subsection{Tree cluster state generation}
We hereby introduce a methodology to generate a tree cluster state with a quantum metasurface, highlighting a basic example of a binary tree with three levels (illustrated in Figure \ref{tree}b). The tree cluster states are useful for fault-tolerant quantum communications \cite{borregaard_one-way_2020}.
To generate a scalable tree cluster protocol we require an implementation of an inheritance gate, symbolized as E, in conjunction with a CZ gate and a Hadamard gate delineated in \cite{zhan_deterministic_2020}. 

The simple tree cluster we will examine can be generated by implementing the circuit illustrated in Figure \ref{tree}c. It consists of two 1D clusters, each comprising three qubits (labeled as 1,5,2 and 3,6,4 in Figure \ref{tree}b). The parent nodes of these clusters (labeled as 5 and 6 in Figure \ref{tree}b) are entangled with the ancilla using a CZ gate. Subsequently, the state of the ancilla is transferred to a photon via an E gate (as detailed below). This procedure enables the generation of an entire tree composed of photonic qubits, which can be transmitted by optics for further use. A comprehensive diagram of this protocol can be found in Figure \ref{tree}a,c, and a detailed proof of the target state generation is provided in the Appendix (see Appendix \ref{supplement}).

\label{E}
To introduce the E gate, we draw inspiration from the approach presented in \cite{zhan_deterministic_2020}. This gate is a two-qubit operation designed to transfer the state of one qubit to a second qubit while leaving the first qubit in a specific non-superpositioned state. Formally, it transfers the state
\begin{equation}
    \left(\alpha\left|g\right\rangle _{a}\left|\psi_{1}\right\rangle _{r}+\beta\left|r\right\rangle _{a}\left|\psi_{0}\right\rangle _{r}\right)\left|1\right\rangle _{p}
\end{equation} 
to \begin{equation}
\left|g\right\rangle _{a}\left(\alpha\left|\psi_{1}\right\rangle _{r}\left|1\right\rangle _{p}+\beta\left|\psi_{0}\right\rangle _{r}\left|0\right\rangle _{p}\right)
\end{equation} 
where $a$ denotes the QM ancillary atom, $p$ signifies the photon, and $r$ represents the remaining part of the state.

The E gate serves the purpose of transferring the state of the ancillary atom to a photon. This enables entanglement of a photon with the ancilla, subsequently transferring the ancilla's state to this photon. The implementation of this gate involves utilizing a CNOT gate between the ancilla (control) and a photon (target), as described above to transfer \((\alpha\left|g\right\rangle _{a}\left|\psi_{1}\right\rangle _{r}+\beta\left|r\right\rangle _{a}\left|\psi_{0}\right\rangle _{r})\left|1\right\rangle _{p}\)  to  \(\alpha\left|g\right\rangle _{a}\left|\psi_{1}\right\rangle _{r}\left|1\right\rangle _{p}+\beta\left|r\right\rangle _{a}\left|\psi_{0}\right\rangle _{r}\left|0\right\rangle _{p}\). By applying a $\pi$ pulse ($\left|r\right\rangle _{a}\leftrightarrow\left|e\right\rangle _{a}$ transition) and subsequently allowing the atom to decay from $\left|e\right\rangle _{a}$ to $\left|g\right\rangle _{a}$ (see Figure \ref{tree}d), the desired state \(\left|g\right\rangle _{a}(\alpha\left|\psi_{0}\right\rangle _{r}\left|1\right\rangle _{p}+\beta\left|\psi_{1}\right\rangle _{r}\left|0\right\rangle _{p})\) is obtained.

Overall, we generate the tree cluster state by first creating two linear cluster branches using sequences of Hadamard and CNOT gates, entangling the ancilla atom with the corresponding photonic qubits. The middle nodes of each branch are then entangled with the ancilla via additional CZ gates. Finally, the E gate transfers the quantum state of the ancilla to a new photonic qubit through a CNOT operation followed by a decay, projecting the ancilla into
\(\left|g\right\rangle _{a}\) while completing the structure of the tree. This process creates a tree cluster state composed entirely of photonic qubits. These can be efficiently distributed through photonic channels for various applications.

An additional advantage of implementing our protocols with a quantum metasurface is its inherent parallelism: By exploiting the spatial degrees of freedom, multiple ancilla–photon interactions can occur simultaneously at different sites on the metasurface using multiple optical paths. This enables the concurrent generation of cluster branches or subgraphs within a single operational cycle, as a potential speed-up mechanism and throughput of fast large-scale photonic cluster state generation. 

\begin{figure*}[ht!]
\centering
\centering\includegraphics[width=0.9\textwidth,height=0.6\textheight]{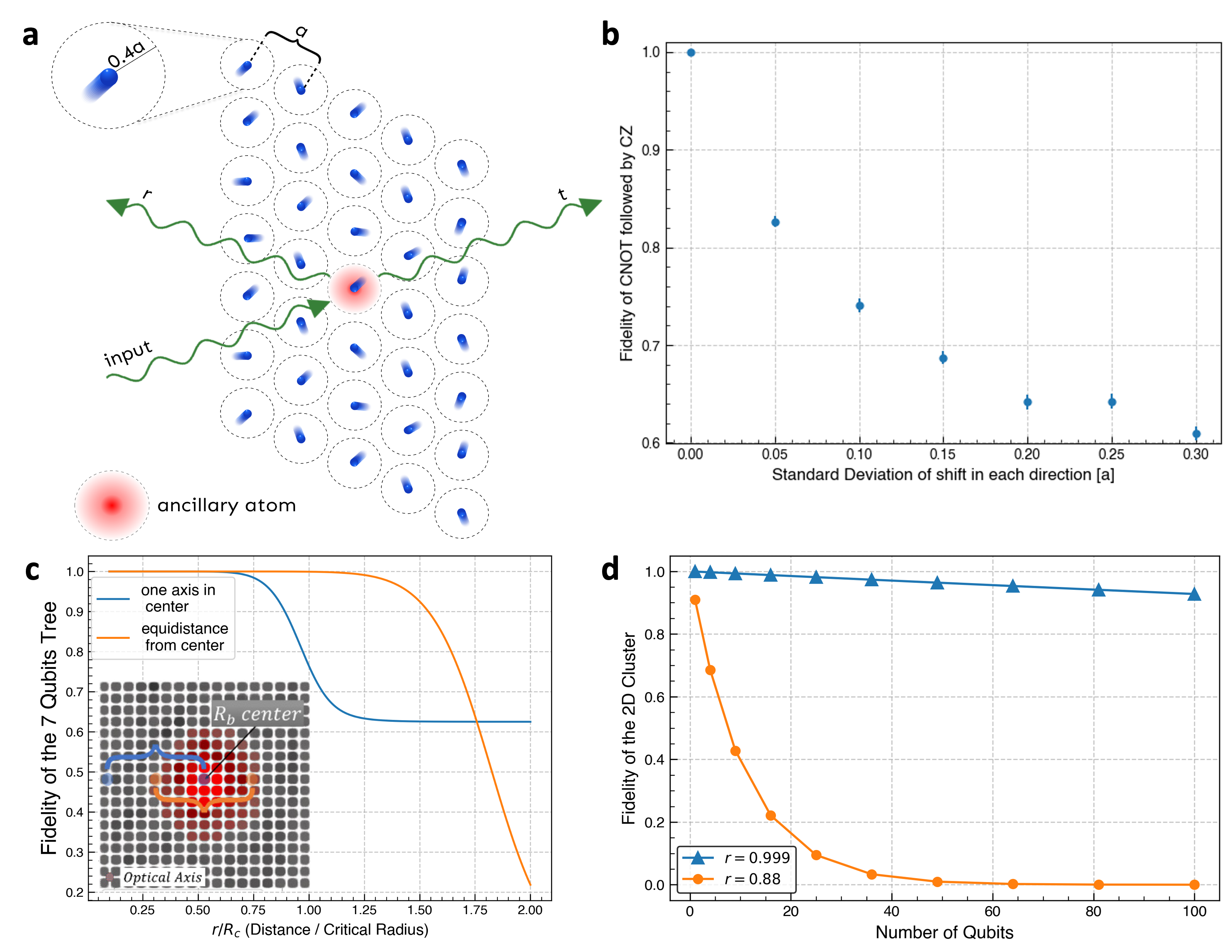}
\caption{\label{deviation} \textbf{a} A quantum metasurface under fluctuations in atomic location. The s.d. of the movement is $0.4a$ in this illustration. As the temperature increases, thermal fluctuations induce movements in the atoms, leading to imperfections in the array. To investigate this, we conducted simulations to assess the loss of reflectivity under different random atomic movements. \textbf{b} Fidelity of optical path 2, defined as a CNOT gate followed by a CZ gate, as specified in the protocol. \textbf{c} Fidelity decay of the tree state as a function of the separation between the two optical paths, due to the finite Rydberg Blockade effect. Two configurations are examined: one with both optical axes equidistant from the center of the Rydberg Blockade, and another with optical path 1 is at the center and optical path 2 offset by a specified distance. \textbf{d} Fidelity of the 2D cluster state versus state size for two scenarios: one with minimal positional disorder corresponding to near-ideal reflection coefficient (0.99), and the other with a positional standard deviation of $0.05a$($0.01\lambda$), which corresponds to a simulated reflection coefficient of 0.88.}
\end{figure*}

\section{Fidelity analysis for practical implementation systems}
\label{Fidelity}

One significant challenge in our protocols arises from the loss mechanism occurring when a photon undergoes imperfect reflection due to factors such as atomic thermal fluctuations. For example, in practical experimental realizations that evolve Rubidium atoms trapped in photonic lattices, a main imperfection occurs in atom location as a result of the atoms' thermal motion \cite{srakaew_subwavelength_2023, rui_subradiant_2020}. 
To obtain realistic values for the fidelities for practical experimental consideration, we introduced imperfections into the system by simulating random displacements of the atoms (see Figure \ref{deviation}a), with the extent of these displacements governed by a specified standard deviation. 

We conducted numerical analysis to estimate the response to light of such systems, to find the practical reflectivity that enters the protocols' fidelity. We simulated the interaction of a Gaussian beam with a wavelength of $0.7\mu m$ with a two-dimensional metasurface composed of dipoles. The metasurface was modeled as a square lattice of $20\times20$ scatterers with atom separations of $0.21\lambda$, where positional disorder was introduced via a Gaussian perturbations with different parameters. Using a dyadic Green’s function formalism \cite{novotny_principles_2012, shahmoon_cooperative_2017}, we numerically computed the multiple scattering between dipoles and solved for the resulting electric field. The incident field is a linearly polarized Gaussian beam, and the total scattered field is calculated for a range of disorder strengths. For each disorder realization, we calculated the reflection amplitude in the vicinity of the array following the scattering event, and evaluated the reflectivity amplitude of the scattered field along the optical axis as a function of propagation distance. The results are presented in Figures \ref{deviation} and \ref{fidelity_graph}. The protocol for the tree cluster state requires two different optical paths: one (path 1, from Figure \ref{tree}a) that is not affected and is assumed to be in an ideal location with respect to the ancilla, and the other (path 2) that suffers from errors. The imperfect fidelity of path 2 affects only specific qubits out of the full tree state we analyze, hence the overall fidelity does not decay to zero. We used the approach detailed in appendix \ref{finite_rydberg} to estimate the error resulting from the reduction in Rydberg interaction with distance (see Figure \ref{deviation}c). For the 2D cluster, each photon undergoes the same CNOT and CZ process. Therefore, the total fidelity scales with the number of photons as displayed in Figure \ref{deviation}d.

To ensure statistical robustness, we conducted 100 simulation runs, systematically varying the random displacements and averaging the results over hundreds of iterations. From these numerical simulations, we obtained statistically robust estimates of the reflectivity and angular distribution as functions of disorder, and evaluated the reflection coefficient for various atom spacings within the QM. A detailed convergence analysis is presented in Appendix Figure \ref{convergance}, providing insights into the stability and convergence behavior of our simulations.

For example, the fidelity of the tree from Figure \ref{tree}b is:

\[
\mathcal{F}_{n=7}(r)
= \left\lvert
\begin{aligned}
\frac{1}{2^{7/2}\sqrt{N_r}}&\Bigl[\langle 0|_{a}\langle \psi_{0}|^{\otimes 2}
     + \langle 1|_{a}\langle \psi_{1}|^{\otimes 2}\Bigr] \\
&\quad\times \Bigl[ |0\rangle_{a}|\phi_{0}\rangle^{\otimes 2}
     + |1\rangle_{a}|\phi_{1}\rangle^{\otimes 2}\Bigr]
\end{aligned}
\right\rvert^{2}
\]

where
\begin{align*}
|\psi_{0}\rangle &= |+0+\rangle + |-1-\rangle, \\
|\psi_{1}\rangle &= |+0+\rangle - |-1-\rangle, \\
|\phi_{0}\rangle &= |00+\rangle + r|10+\rangle + r|01-\rangle - r^{2}|11-\rangle, \\
|\phi_{1}\rangle &= |00+\rangle + r|10+\rangle - r^{2}|01-\rangle + r^{3}|11-\rangle, \\
N_r &= 2 + 10|r|^2 + 23|r|^4 + 32|r|^6 \\
        &\quad+ 29|r|^8 + 18|r|^{10} + 9|r|^{12} + 4|r|^{14} + |r|^{16}.
\end{align*}
and r is the reflectance coefficient due to imperfections in the QM, and the normalization factor is $N_r$. In total, we get 
\begin{align*}
\mathcal{F}_{n=7}(r) 
= \frac{1}{2^{7} N_r} \;\Bigg|\, 
&2 + 10r + 23r^2 + 32r^3 \\
&+ 29r^4 + 18r^5 + 9r^6 + 4r^7 + r^8
\Bigg|^{2}
\end{align*}
For a reflection coefficient of $r=0.999$, the resulting fidelity is 0.999, whereas for $r=0.88$ the fidelity is reduced to 0.962.

\section{Discussion}
In conclusion, we have proposed an implementation for fundamental building blocks of cluster states, namely the CNOT, CZ, E, and Hadamard gates. These gates serve as crucial components for constructing various cluster states. The sequential implementation of fundamental quantum gates through the ancilla-mediated interactions provides a scalable framework for creating complex multi-qubit entangled states by employing many sequential photonic qubits. 

In particular, we have presented two specific cluster states: a 2D cluster state suitable for quantum computation and a tree cluster state applicable for data encryption and transmission. By leveraging the properties of the Quantum Metasurface, we have demonstrated the physical realization of these gates, which enables the creation of diverse cluster states. The 2D cluster state holds promise for quantum computation, while the tree cluster state offers potential applications in secure data transfer. Importantly, The combination of our analysis along with the proposed protocols in \cite{antman_ron_atom-atom_2024} offers a scalable atom-photon entanglement scheme. Our work contributes to advancing quantum information processing by providing feasible protocols for constructing cluster states using the proposed gate implementations.

\begin{acknowledgments}
This research was funded by Grant No. 2021775 from the United States-Israel Binational Science Foundation (BSF) and by Grant No. 2207972. From the United States National Science Foundation (NSF) and ISF Grant No. 2402/22. This work has received funding from the European Research Council (ERC) under the European Union's Horizon 2020 research and innovation programme (Grant Agreement No. 101117845).
\end{acknowledgments}

\appendix
\section{Creation and Verification of the Tree Cluster}
\label{supplement}

\renewcommand{\thefigure}{\Alph{section}\arabic{figure}}
\setcounter{figure}{0}
\begin{figure*}[ht!]
\centering
\centering\includegraphics[width=0.9\textwidth]{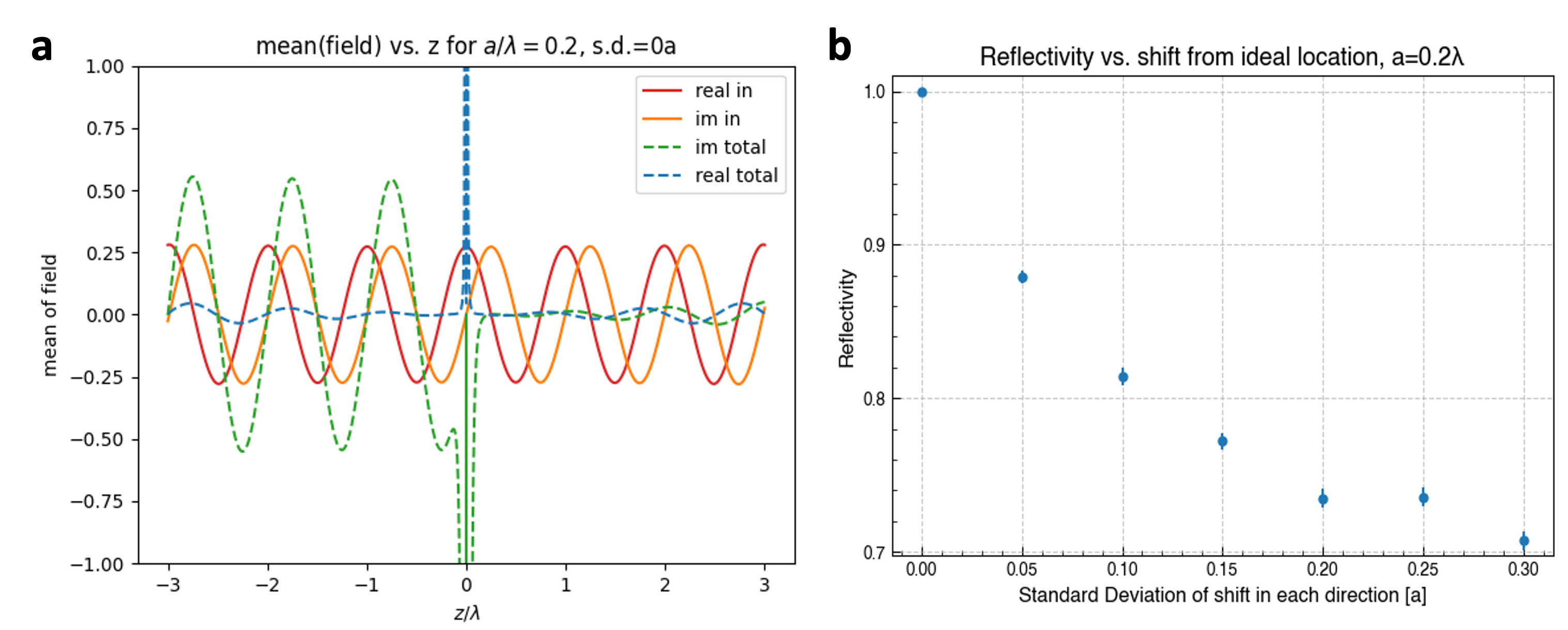}
\caption{\label{fidelity_graph} 
\textbf{a} The simulated electromagnetic field for zero movement. There is a destructive interference behind the QM and instructive interference of the imaginary part when reflected. 
\textbf{b} Our results demonstrate a decrease in the reflectivity of the Quantum Metasurface as the average distance between the atoms deviates from their ideal positions due to thermal effects.}
\end{figure*}
\begin{figure*}[ht!]
\centering
\centering\includegraphics[width=0.9\textwidth]{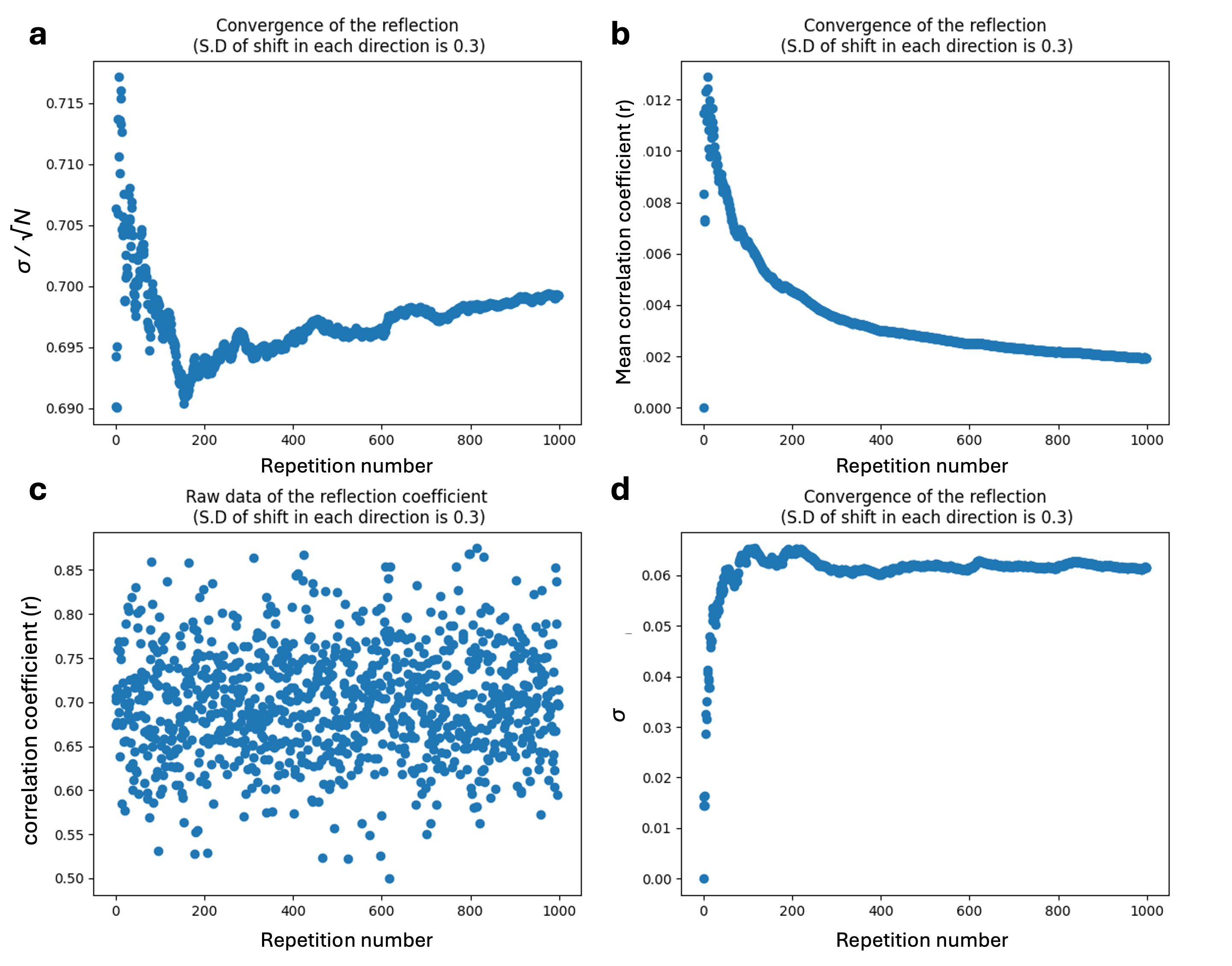}
\caption{\label{convergance} Convergence analysis of the simulated reflection using 1,000 runs. 
\textbf{a} Standard deviation divided by $\sqrt{N}$, with $N$ the number of runs, of the reflection coefficient as a function of the repetition number. 
\textbf{b} Mean reflection coefficient as a function of the repetition number. 
\textbf{c} Reflection coefficient of each run. 
\textbf{d} Standard deviation of the reflection coefficient as a function of the repetition number.}
\end{figure*}

We demonstrate that our protocol successfully generates the desired tree cluster state by constructing it from quantum gates and photons, and show that it reproduces the expected theoretical expressions. Furthermore, we outline a verification method based on the measurement of the corresponding stabilizer operators. 
\subsection{1D Cluster State Creation Using Protocol of CNOT and Hadamard Gates}
Starting with 4 qubits in the $\left|0\right\rangle$ state — 3 photons and an ancilla — we apply a Hadamard gate on the ancilla, and then CNOT gates from the ancilla to each photon:
\[\prod_{i=1}^{3} H_{{a}} \, \text{CNOT}_{{a},i}\left(H_{{a}}\left|{{0}}000\right\rangle\right) = \prod_{i=1}^{3} H_{{a}} \, \text{CNOT}_{{a},i} \left|{{+}}000\right\rangle\]
Applying these gates yields:

\begin{equation}
\begin{aligned}
& \frac{1}{4} \Big( 
   \left| {0}000 \right\rangle + \left| {1}000 \right\rangle 
   + \left| {0}010 \right\rangle + \left| {1}010 \right\rangle 
\Big) \\
& + \frac{1}{4} \Big(
   \left| {0}100 \right\rangle + \left| {1}100 \right\rangle 
   - \left| {0}110 \right\rangle - \left| {1}110 \right\rangle 
\Big) \\
& + \frac{1}{4} \Big(
   \left| {0}001 \right\rangle - \left| {1}001 \right\rangle 
   - \left| {0}011 \right\rangle + \left| {1}011 \right\rangle 
\Big) \\
& + \frac{1}{4} \Big(
   \left| {0}101 \right\rangle - \left| {1}101 \right\rangle 
   + \left| {0}111 \right\rangle - \left| {1}111 \right\rangle 
\Big)
\end{aligned}
\end{equation}

Grouping terms by measurement result:
If we measure the atom in the computational basis, the resulted states for the two possible atomic state outcomes are:
\begin{equation}
\text{Outcome } 0: \quad 
\frac{1}{\sqrt{8}}
\begin{aligned}[t]
& |000\rangle + |010\rangle + |100\rangle - |110\rangle \\
& + |001\rangle - |011\rangle + |101\rangle + |111\rangle
\end{aligned}
\end{equation}

\begin{equation}
\text{Outcome } 1: \quad 
\frac{1}{\sqrt{8}}
\begin{aligned}[t]
& |000\rangle + |010\rangle + |100\rangle - |110\rangle \\
& - |001\rangle + |011\rangle - |101\rangle - |111\rangle
\end{aligned}
\end{equation}
In either case, the state is a 1D 3-body cluster state. For convenience, we will work with the first outcome, but both outcomes represent the same cluster state up to local transformations. A more compact expression is:
\begin{equation}
\frac{1}{\sqrt{8}}
\begin{aligned}[t]
& \Big( |000\rangle + |001\rangle + |100\rangle + |101\rangle \\
& \quad + |010\rangle - |011\rangle - |110\rangle + |111\rangle \Big) \\
&= \frac{1}{2} \Big( |00+\rangle + |10+\rangle + |01-\rangle - |11-\rangle \Big) \\
&= \frac{1}{\sqrt{2}} \Big( |+0+\rangle + |-1-\rangle \Big)
\end{aligned}
\end{equation}
As in the protocol, we use two such clusters and entangle them via a controlled-$Z$ (CZ) gate between the ancilla and photons 5 and 6 (as in Fig. \ref{tree}b):
\[\text{CZ}_{{a},5-6}\left[\frac{1}{2^{3/2}}\left(\left|+\right\rangle_{{a}}\left(|+0+\rangle+|-1-\rangle\right)^{\otimes2}\right)\right]\]
By expanding the above:
\begin{align}
= \text{CZ}_{{a},5-6} \Bigg[
\frac{1}{2^{3/2}} &\Big(
\left|0\right\rangle_{{a}} \left(|+0+\rangle + |-1-\rangle\right)^{\otimes 2} \notag \\
&\quad + \left|1\right\rangle_{{a}} \left(|+0+\rangle + |-1-\rangle\right)^{\otimes 2}
\Big) \Bigg] \notag \\
= \frac{1}{2^{3/2}} &\Big(
\left|0\right\rangle_{{a}} \left(|+0+\rangle + |-1-\rangle\right)^{\otimes 2} \notag \\
&\quad + \left|1\right\rangle_{{a}} \left(|+0+\rangle - |-1-\rangle\right)^{\otimes 2}
\Big)
\end{align}
which is identical to the expected entangled cluster state (see e.g., \cite{zhan_deterministic_2020, borregaard_one-way_2020}) if we relabel the photons appropriately (specifically swapping photon indices, corresponding to $\left|7152364\right\rangle$ in Figure~\ref{tree}b).

In the case where the reflection is not perfect and we have a reflection coefficient r, we would get
\begin{equation}
\begin{aligned}
&\frac{1}{\sqrt{2}(1+|r|^2)^{3/2}}
\Big[
(|0 000\rangle + |1 000\rangle + r|0 010\rangle + r|1 010\rangle) \\[3pt]
&\quad + (r|0 100\rangle + r|1 100\rangle - r^2|0 110\rangle - r^2|1 110\rangle) \\[3pt]
&\quad + (|0 001\rangle - |1 001\rangle - r|0 011\rangle + r|1 011\rangle) \\[3pt]
&\quad + (r|0 101\rangle - r|1 101\rangle + r^2|0 111\rangle - r^2|1 111\rangle)
\Big]
\end{aligned}
\end{equation}›
Grouping terms by measurement result:
If we measure the atom in the computational basis, the resulted states for the two possible atomic state outcomes are:
\begin{align}
\text{Outcome } 0: &\quad \frac{1}{(1+|r|^2)^{3/2}} \Big(
|000\rangle + r|010\rangle + r|100\rangle - r^2|110\rangle \notag \\
&\quad + |001\rangle - r|011\rangle + r|101\rangle + r^2|111\rangle \Big) \\
\text{Outcome } 1: &\quad \frac{1}{(1+|r|^2)^{3/2}} \Big(
|000\rangle + r|010\rangle + r|100\rangle - r^2|110\rangle \notag \\
&\quad - |001\rangle + r|011\rangle - r|101\rangle - r^2|111\rangle \Big)
\end{align}
Without loss of generality we would continue with outcome 0. 
As above, we use two such clusters and entangle them via a CZ gate between the ancilla and photons 5 and 6, and recalling that CZ also uses reflectance of QM, we get:
\begin{align*}
\frac{1}{\sqrt{N_r}} \Big( & 
\left|0\right\rangle_{{a}} \Big(
\left|000\right\rangle + r(\left|100\right\rangle + \left|010\right\rangle + \left|001\right\rangle) \\
&\quad + r^2(-\left|011\right\rangle + \left|101\right\rangle - \left|110\right\rangle) + r^3 \left|111\right\rangle 
\Big)^{\otimes 2} \\
&+ \left|1\right\rangle_{{a}} \Big(
\left|000\right\rangle + r(\left|100\right\rangle - r \left|010\right\rangle + \left|001\right\rangle) \\
&\quad + r^2(r \left|011\right\rangle + \left|101\right\rangle + r \left|110\right\rangle) - r^4 \left|111\right\rangle
\Big)^{\otimes 2} \Big)
\end{align*}
for $N_r=2+10|r|^2+23|r|^4+32|r|^6+29|r|^8+18|r|^{10}+9|r|^{12}+4|r|^{14}+|r|^{16}$
\subsection{Detailed stabilizer verification}
A cluster (graph) state is the unique simultaneous +1 eigenstate of local stabilizer generators. Verification of such a state can be performed by measuring these stabilizer operators-each acting on a given qubit and its nearest neighbors-confirming that the measurement outcomes yield +1.
The stabilizer is given by
\[K_{i} = \sigma_{i}^{x} \bigotimes_{j \in \mathcal{N}(i)} \sigma_{j}^{z}\]
where $\mathcal{N}(i)$ denotes the set of neighboring qubits of $i$ \cite{hein_m_entanglement_2006}.

\noindent\textbf{Verification for qubit 7 (numbers as in Figure \ref{tree}b):}
\begin{widetext}
\begin{align*}
K_{7}&\left(\frac{1}{2^{3/2}}\left(\left|0\right\rangle_{7}\left(|+0+\rangle+|-1-\rangle\right)^{\otimes2} + \left|1\right\rangle_{7}\left(|+0+\rangle-|-1-\rangle\right)^{\otimes2}\right)\right) \\
&= \sigma_{1}^{x} \sigma_{5}^{z} \sigma_{6}^{z} \left(\frac{1}{2^{3/2}}\left|0\right\rangle_{7}\left(|+0+\rangle+|-1-\rangle\right)^{\otimes2} + \left|1\right\rangle_{7}\left(|+0+\rangle-|-1-\rangle\right)^{\otimes2}\right) \\
&= \frac{1}{2^{3/2}}\left(\left|1\right\rangle_{7}\left(|+0+\rangle-|-1-\rangle\right)^{\otimes2} + \left|0\right\rangle_{7}\left(|+0+\rangle+|-1-\rangle\right)^{\otimes2}\right) \\
&= \frac{1}{2^{3/2}}\left(\left|0\right\rangle_{7}\left(|+0+\rangle+|-1-\rangle\right)^{\otimes2} + \left|1\right\rangle_{7}\left(|+0+\rangle-|-1-\rangle\right)^{\otimes2}\right)
\end{align*}
\end{widetext}

\noindent\textbf{Verification for middle qubits (5 and 6):}
We present the verification for qubit 5; the procedure and results apply identically to qubit 6 due to symmetry.
\begin{widetext}
\begin{align*}
K_{5}&\left(\frac{1}{2^{3/2}}\left(\left|0\right\rangle_{7}\left(|+0+\rangle+|-1-\rangle\right)^{\otimes2} + \left|1\right\rangle_{7}\left(|+0+\rangle-|-1-\rangle\right)^{\otimes2}\right)\right) \\
&= \sigma_{1}^{z} \sigma_{2}^{z} \sigma_{7}^{z} \sigma_{5}^{x} \left(\frac{1}{2^{3/2}}\left|0\right\rangle_{7}\left(|+0+\rangle+|-1-\rangle\right)^{\otimes2} + \left|1\right\rangle_{7}\left(|+0+\rangle-|-1-\rangle\right)^{\otimes2}\right) \\
&= \sigma_{1}^{z} \sigma_{2}^{z} \left(\frac{1}{2^{3/2}}\left|0\right\rangle_{7}\left(|+1+\rangle+|-0-\rangle\right)\left(|+0+\rangle+|-1-\rangle\right) - \left|1\right\rangle_{7}\left(|+1+\rangle-|-0-\rangle\right)\left(|+0+\rangle-|-1-\rangle\right)\right) \\
&= \frac{1}{2^{3/2}}\left(\left|0\right\rangle_{7}\left(|+0+\rangle+|-1-\rangle\right)^{\otimes2} + \left|1\right\rangle_{7}\left(|+0+\rangle-|-1-\rangle\right)^{\otimes2}\right)
\end{align*}
\end{widetext}

\noindent\textbf{Verification for leaves (qubits 1, 2, 3, 4):}
We present the verification for qubit 1; the procedure and results apply identically to other qubits due to symmetry.
\begin{widetext}
\begin{align*}
K_{1}&\left(\frac{1}{2^{3/2}}\left(\left|0\right\rangle \left(|+0+\rangle+|-1-\rangle\right)^{\otimes2} + \left|1\right\rangle \left(|+0+\rangle-|-1-\rangle\right)^{\otimes2}\right)\right) \\
&= \sigma_{1}^{x} \sigma_{5}^{z} \left(\frac{1}{2^{3/2}}\left|0\right\rangle \left(|+0+\rangle+|-1-\rangle\right)^{\otimes2} + \left|1\right\rangle \left(|+0+\rangle-|-1-\rangle\right)^{\otimes2}\right) \\
&= \frac{1}{2^{3/2}} \left(\left|0\right\rangle \left(|+0+\rangle+|-1-\rangle\right)^{\otimes2} + \left|1\right\rangle \left(|+0+\rangle-|-1-\rangle\right)^{\otimes2}\right)
\end{align*}
\end{widetext}

\section{decay of fidelity due to finite Rydberg blockade}
\label{finite_rydberg}
When the ancilla is in the Rydberg state, the QM becomes reflective, and the reflectivity coefficient at distance $r_d$ from the ancilla is
\[
1-\frac{1}{-i\left(\frac{R_c} {r_d} \right)^{6} + 1} = \frac{1}{1+i\left(\frac{r_d}{R_c} \right)^{6}},
\quad \]
where $R_c$ is the critical interaction radius, dofined as \[ \quad
R_c = \left( \frac{(\gamma + \Gamma) c_6}{2 |\Omega_p|^{2}} \right)^{1/6}
\]
for $c_6$ the Van der Waals coefficient, $(\gamma + \Gamma)$ the total decay rate (or linewidth) of the excited state, and $\Omega_p$ is the Rabi frequency of the pump laser that couples the excited state to the Rydberg state (see \cite{bekenstein_quantum_2020}).

\bibliographystyle{apsrev4-2}
\bibliography{Quantum}

\end{document}